\documentclass[a4paper,11pt]{article}

\usepackage{textcomp}
\usepackage{amssymb}
\usepackage{units}

\usepackage[overload]{textcase}

\usepackage{graphicx}
\usepackage{subfigure}

\usepackage[labelsep=period]{caption}

\usepackage{amssymb}

\usepackage{amsmath}
\usepackage{mathabx}

\usepackage[sort&compress]{natbib}
\bibpunct{[}{]}{,}{n}{}{}

\begin{document}

\thispagestyle{plain}
\setcounter{page}{1}

\vfill

\begin{center}

   \vspace{0.3cm}
   {\Large \textbf{Dynamics of a sphere with inhomogeneous slip boundary conditions in Stokes flow}} \\
   \vspace{0.3cm}
   {\large Geoff Willmott}\\
   {\large \emph{Industrial Research Limited, 69 Gracefield Rd, Lower Hutt, New Zealand}}\\
	\noindent{(Received 11 January 2008)}\\
	\vspace{0.3cm}
	{\large Email: g.willmott@irl.cri.nz}\\
	\vspace{0.3cm}
    {\large Phone: (64) (0)4 931 3220}\\
    \vspace{0.3cm}
    {\large Fax: (64) (0)4 931 3117}\\

\renewcommand{\abstractname}{}

\begin{abstract}
\noindent The dynamic resistance of a sphere with a general inhomogeneous slip boundary condition is analysed in Newtonian unbounded uniform flow at low Reynolds number. The boundary condition is treated as a perturbation to a homogeneous sphere, assuming that the slip length magnitude $b$ is much smaller than the sphere radius $a$. To first order, the effect of inhomogeneous slip is the same as that of a radial deformity of magnitude $b$. Full resistance tensors are presented and the dynamics of a hemispherical inhomogeneous sphere, such as a Janus particle, are explicitly calculated.   
\end{abstract}
\end{center}

\vspace{0.3cm}
The motion of micro- and nanospheres in fluids is of increasing importance in several emerging fields of research. Spheres have attracted interest for biomedical applications such as targeting cancer cells \cite{475, 496}. Magnetic microspheres have frequently been used in recent research \cite{494}, while small particles have the potential to play a significant role in microfluidic and MEMS technologies. Fabrication of micro- and nanospheres is widespread, because it is relatively easy to obtain spherical particles with homogeneous surface chemistry due to minimization of surface tension. Inhomogeneous spherical particles such as dual-hemispherical Janus particles \cite{491} (including magnetic Janus particles \cite{501}) are also being fabricated. 

Manipulation of small particles in fluids has been achieved by various means, particularly using electromagnetic methods, which have been described analytically \cite{495}. Other factors relevant to particle manipulation that are less well studied include orientation dependence \cite{475}, particle inhomogeneity and surface slip. Slip occurs when there is non-zero relative motion between a solid surface and the fluid immediately adjacent to that surface, violating the non-slip boundary condition (NSBC). The presence of slip on a particle surface alters the shear forces on that surface, which are otherwise very stable at low Reynolds number, thereby introducing novel manipulation properties that do not require application of external fields. The rotational dynamics are likely to be of intrinsic importance for inhomogeneous and Janus particles. In the last 15 years, there has been great interest in practically achieving surface slip in microfluidic flows \cite{335, 303}. Experimental work has shown that, for a Newtonian fluid at a smooth, hydrophobic surface, non-zero slip does occur and can be greatly intensified using surfaces engineered on the microscale \cite{437, 436, 434, 435, 498}. Complementary theoretical work has most frequently used Navier's formulation of the slip boundary condition \cite{335}.  

Analysis of spherical and near-spherical particles in fluid flow at low Reynolds numbers has consistently been of interest over the past two hundred years. Happel and Brenner \cite{463} produced a summary of relevant analyses, covering the flows surrounding particles and the resulting dynamics. In particular, they presented the solution for a sphere with a homogeneous slip boundary condition. The problem of a slightly deformed sphere with a homogeneous slip boundary condition has since been solved approximately \cite{479, 492}.

In this Letter, the dynamic response of an impermeable spherical particle with a spatially varying, inhomogeneous slip boundary condition is analysed for the first time in streaming Newtonian flow at low Reynolds number. Slip is formulated as a general function of surface area and applied as a perturbation to the solution for a homogeneous NSBC sphere. This approach gives first-order results for the resistance tensor of the particle for any slip boundary condition and any relative flow direction, providing the slip length magnitude is much smaller than the sphere's radius. Results are calculated for some specific inhomogeneous boundary conditions, including those relevant for Janus particles.  

The equation of motion for steady flow of an incompressible fluid ($\nabla.\mathbf{v}=0$) at low Reynolds number is Stokes' equation, 

\begin{equation}\label{eq:Navier-Stokes}
\eta\nabla^2\mathbf{v}=\nabla p,\\
\end{equation}

\noindent where $\mathbf{v}$ is the velocity field for a Newtonian fluid of density $\rho$ and viscosity $\eta$, and the pressure term $p$ absorbs any body forces. Solutions to Stokes' equation are dependent on flow geometry and boundary conditions. There is a general methodology for finding $\mathbf{v}$ for unbounded uniform flow external to a particle in viscous flow, and the force and torque on a spherical particle \cite{463}. For inhomogeneous boundary conditions, it is not practical to analytically find the full solution, which requires several expansions of spherical harmonic functions.  

Navier's slip boundary condition \cite{335} states that the component of fluid velocity tangent to a solid surface ($v_\parallel$) is proportional to the shear rate at the surface, 

\begin{equation}\label{eq:Slip defn}
v_\parallel=b\mathbf{n}\cdot\left(\nabla\mathbf{v}+\left(\nabla\mathbf{v}\right)^T\right)\cdot\left(\mathbf{I}-\mathbf{nn}\right).
\end{equation}

\noindent Here, $\mathbf{n}$ is the unit normal to the surface and the equation is valid in the rest frame of the wall \cite{455}. The parameter $b$ has units of length and is often referred to as the \textquoteleft slip length'. For an impermeable, homogeneous surface, Eq.~\ref{eq:Slip defn} reduces to $v_\parallel=b_\parallel v_\parallel'$, where $v_\parallel'$ is the spatial gradient of $v_\parallel$ normal to the solid surface and $b_\parallel=b$ if the solid-liquid interface is assumed to be planar, as has been the case in most recent slip studies. The derivation of $b$ for a curved surface \cite{455} yields

\begin{equation}\label{eq:curved slip defn}
\frac{1}{b_\parallel}=\frac{1}{b}-\frac{1}{R},
\end{equation}

\noindent where $R$ is the radius of curvature of the interface, defined as positive for a convex solid surface. Note that $b_\parallel=0$ when $b=0$; $b_\parallel\approx b$ when $b\ll R$; and if $R$ and $b$ are of comparable magnitude, $b_\parallel>b$. 

In order to describe an inhomogeneous boundary condition on the surface of a sphere, the slip length $bf\left(\theta,\phi\right)$ is used, where $\theta$ and $\phi$ are spherical polar coordinates and $f\left(\theta,\phi\right)$ is a function of $O(1)$. When $b$ is small compared with sphere radius $a$, a perturbed flow around the sphere can be expanded in powers of $\epsilon=-(b_\parallel$/$a)$, where $\lvert\epsilon\rvert\ll 1$ and therefore $b_\parallel\approx b$:  

\begin{equation}\label{eq:Pert1}
\mathbf{v}=\mathbf{v}^{(0)}+\epsilon\mathbf{v}^{(1)}+\epsilon^2\mathbf{v}^{(2)}+...
\end{equation}

\noindent The inhomogeneous version of the flat-surface slip boundary condition (Eq.~\ref{eq:Slip defn}) is

\begin{equation}\label{eq:Pert2}
\mathbf{v}
=-\epsilon a f\left(\theta,\phi\right)\frac{\partial \mathbf{v}}{\partial r},
\end{equation}

\noindent where $f\left(\theta,\phi\right)$ can be expanded in terms of surface spherical harmonics $f_k\left(\theta, \phi\right)$, such that $f\left(\theta,\phi\right)=\sum_{k=0}^\infty f_k\left(\theta, \phi\right)$. Comparing Eqs.~\ref{eq:Pert1} and \ref{eq:Pert2}, the first two terms in the expansion of $\mathbf{v}$ give the boundary conditions,

\begin{equation}\label{eq:PertBC1}
\begin{split}
\mathbf{v}^{(0)}\left(r=a, \theta, \phi\right)&=\mathbf{0},\\
\mathbf{v}^{(1)}\left(r=a, \theta, \phi\right)&=-af\left(\theta,\phi\right)\frac{\partial \mathbf{v}^{(0)}}{\partial r}.
\end{split}
\end{equation}

\noindent For incident unbounded uniform flow characterised by the velocity vector $\mathbf{U}$, the boundary conditions far from the sphere are

\begin{equation}\label{eq:PertBC2}
\begin{split}
\mathbf{v}^{(0)}\left(r=\infty, \theta, \phi\right)&=\mathbf{U},\\
\mathbf{v}^{(1)}\left(r=\infty, \theta, \phi\right)&=\mathbf{0}.
\end{split}
\end{equation}

\noindent The solution for the zeroth-order field $\mathbf{v}^{(0)}$ is the usual solution for a homogeneous sphere, 

\begin{equation}\label{eq:A4}
\mathbf{v}^{(0)}=\mathbf{U}\cdot\left[\mathbf{I}-\frac{3a}{4}\left(\frac{\mathbf{I}}{r}+\frac{\mathbf{r}\mathbf{r}}{r^3}\right)-\frac{a^3}{4}\left(\frac{\mathbf{I}}{r^3}-3\frac{\mathbf{r}\mathbf{r}}{r^5}\right)\right].
\end{equation}

\noindent The corresponding pressure, drag and torque are similarly well known. To find $\mathbf{v}^{(1)}$, a spherical harmonic expansion is used, 
  
\begin{equation}\label{eq:Pertv1}
\mathbf{v}^{(1)}=\sum^\infty_{k=0}\mathbf{v}_k^{(1)},
\end{equation}

\noindent where each $\mathbf{v}_k^{(1)}$ ($k=0,1,2,...$) satisfies Stokes' equations and (from Eqs.~\ref{eq:PertBC1}, \ref{eq:PertBC2} and \ref{eq:A4}) the boundary conditions

\begin{equation}\label{eq:Pertv1BC2}
\begin{split}
\mathbf{v}_k^{(1)}\left(r=a,\theta,\phi\right)&=-\frac{3}{2}\mathbf{U}\cdot\left(\mathbf{I}-\frac{\mathbf{r}\mathbf{r}}{r^2}\right)f_k\left(\theta,\phi\right),\\
\mathbf{v}_k^{(1)}\left(r=\infty,\theta,\phi\right)&=\mathbf{0}.
\end{split}
\end{equation}

\noindent The requirement that $v_r\left(r=a,\theta,\phi\right)=0$ was not explicitly stated when determining the boundary conditions, but is met by Eqs.~\ref{eq:PertBC2} and \ref{eq:Pertv1BC2}. 

For a full description of the spherical harmonic formulation of $\mathbf{v}^{(1)}$ and the relating dynamics, the reader is referred to the analogous treatment of a slightly deformed sphere given by Happel and Brenner \cite{463}, in which the sphere's surface takes the locus of points  

\begin{equation}\label{eq:Def}
r=a\left(1+ef\left(\theta, \phi\right)\right).
\end{equation}

\noindent When this surface is treated as a perturbation to a sphere under the NSBC, the formulation is coincident with the first-order problem defined by Eqs.~\ref{eq:Navier-Stokes}, \ref{eq:PertBC1} and \ref{eq:PertBC2}, with $e$ being replaced by $\epsilon$. To first order in $\epsilon$, the force and torque on the inhomogeneous slip sphere are 

\begin{equation}\label{eq:PertForce}
\begin{split}
\mathbf{F}&=\mathbf{F}^{(0)}+\epsilon\mathbf{F}^{(1)}+O\left(\epsilon^2\right)\\
&=6\pi\eta a\mathbf{U}-\frac{b_\parallel}{a}6\pi\eta a\left(\mathbf{U}f_0-\frac{1}{10}\left(\mathbf{U}\cdot\nabla\right)\nabla\left(r^2f_2\right)\right)+O\left(\epsilon^2\right),
\end{split}
\end{equation}

\begin{equation}\label{eq:PertTorque}
\begin{split}
\mathbf{T}&=\mathbf{T}^{(0)}+\epsilon\mathbf{T}^{(1)}+O\left(\epsilon^2\right)\\
&=\mathbf{0}+\frac{b_\parallel}{a}6\pi\eta a^2\mathbf{U}\times\nabla\left(rf_1\right)+O\left(\epsilon^2\right).
\end{split}
\end{equation}

The dynamics for a rotating, slightly deformed sphere can be calculated following a similar method, once the boundary conditions are carefully checked. Again the perturbation of the velocity field (Eq.~\ref{eq:Pert1}) is used, where $\lvert\epsilon\rvert\ll 1$. The inhomogeneous slip boundary condition is

\begin{equation}\label{eq:Rot2a}
\mathbf{v}=\boldsymbol{\omega}\times\mathbf{r}-\epsilon af\left(\theta,\phi\right)\frac{\partial \mathbf{v}}{\partial r}.
\end{equation}

\noindent Comparing Eqs.~\ref{eq:Pert1} and \ref{eq:Rot2a}, the boundary conditions at the sphere's surface are

\begin{equation}\label{eq:RotBC1}
\begin{split}
\mathbf{v}^{(0)}\left(r=a,\theta,\phi\right)&=\boldsymbol{\omega}\times\mathbf{r}\\
\mathbf{v}^{(1)}\left(r=a,\theta,\phi\right)&=-af\left(\theta,\phi\right)\frac{\partial \mathbf{v}^{(0)}}{\partial r}.
\end{split}
\end{equation}

\noindent The boundary conditions at $r=\infty$ are $\mathbf{v}^{(0)}=\mathbf{v}^{(1)}=\mathbf{0}$. These boundary conditions are also analogous to the spherical harmonic treatment of a slightly deformed sphere \cite{463}. Conversely to the case for translational motion, only odd harmonics generate a force on the rotating particle, while only even harmonics generate torque. To first order in $\epsilon$, 

\begin{equation}\label{eq:RotForce}
\mathbf{F}=\mathbf{0}+\left(\frac{b_\parallel}{a}\right)6\pi\eta a^2\boldsymbol{\omega}\times\nabla\left(rf_1\right)+O\left(\epsilon^2\right),
\end{equation}

\begin{equation}\label{eq:RotTorque}
\mathbf{T}=-8\pi\eta a^3\boldsymbol{\omega}+\left(\frac{b_\parallel}{a}\right)24\pi\eta a^3\left(\boldsymbol{\omega}f_0-\frac{1}{10}\left(\boldsymbol{\omega}\cdot\nabla\right)\nabla\left(r^2f_2\right)\right)+O\left(\epsilon^2\right).
\end{equation}

Stokes' equations and the boundary conditions are linear, so solutions for translation and rotation can be superposed to give the full resistance tensors \cite{479},

\begin{equation}\label{eq:Resistance}
\begin{pmatrix}
\mathbf{F}^{(i)}\\
\mathbf{T}^{(i)}
\end{pmatrix}=\begin{pmatrix}
6\pi\eta a\mathbf{\Phi}^{(i)}& 6\pi\eta a^2\mathbf{D}^{(i)}\\
6\pi\eta a^2\mathbf{C}^{(i)}& 8\pi\eta a^3\mathbf{\Omega}^{(i)}
\end{pmatrix}\cdot\begin{pmatrix}
\mathbf{U}\\
\boldsymbol{\omega}
\end{pmatrix}.
\end{equation}

\noindent The first-order tensors about the centre of an inhomogeneous, slipping sphere are 

\begin{equation}\label{eq:Dyad1}
\Phi_{ij}^{(1)}=\delta_{ij}f_0-\frac{1}{10}\nabla_i\nabla_j\left(r^2f_2\right),
\end{equation}

\begin{equation}\label{eq:Dyad2}
C_{ij}^{(1)}=D_{ij}^{(1)}=-\varepsilon_{ijk}\nabla_k\left(rf_1\right),
\end{equation}

\begin{equation}\label{eq:Dyad4}
\Omega_{ij}^{(1)}=-3\left(\delta_{ij}f_0-\frac{1}{10}\nabla_i\nabla_j\left(r^2f_2\right)\right),
\end{equation}

\noindent where $\delta_{ij}$ is the Kronecker delta, $\varepsilon_{ijk}$ is the Levi-Civita permutation symbol, and tensor components are given in a Cartesian frame. 

As noted above, the flow field and resulting dynamics for a slip length $b$ are dynamically equivalent to surface deformation $-ea$ for a sphere in viscous flow when $\epsilon\ll 1$ and $e\ll 1$. The mathematical equivalence of these approaches does not mean that they describe physically similar phenomena. The boundary conditions are coincident to first order, but not to second order, because the slip boundary conditions for $\mathbf{v}^{(2)}$ and higher-order terms do not contain second and higher derivatives of the flow profile. When higher-order terms are considered, the difference between $b$ and $b_\parallel$ becomes important. Another physical difference is that surface deformation is independent of strain rate, whereas the rate dependence of $b$ (and, therefore, the linearity of Eq.~\ref{eq:Slip defn}) is an unresolved topic of active research \cite{335}.  

\begin{figure}
\begin{center}
\subfigure[]{\label{Fig:Coordinates}\includegraphics[width=4cm]{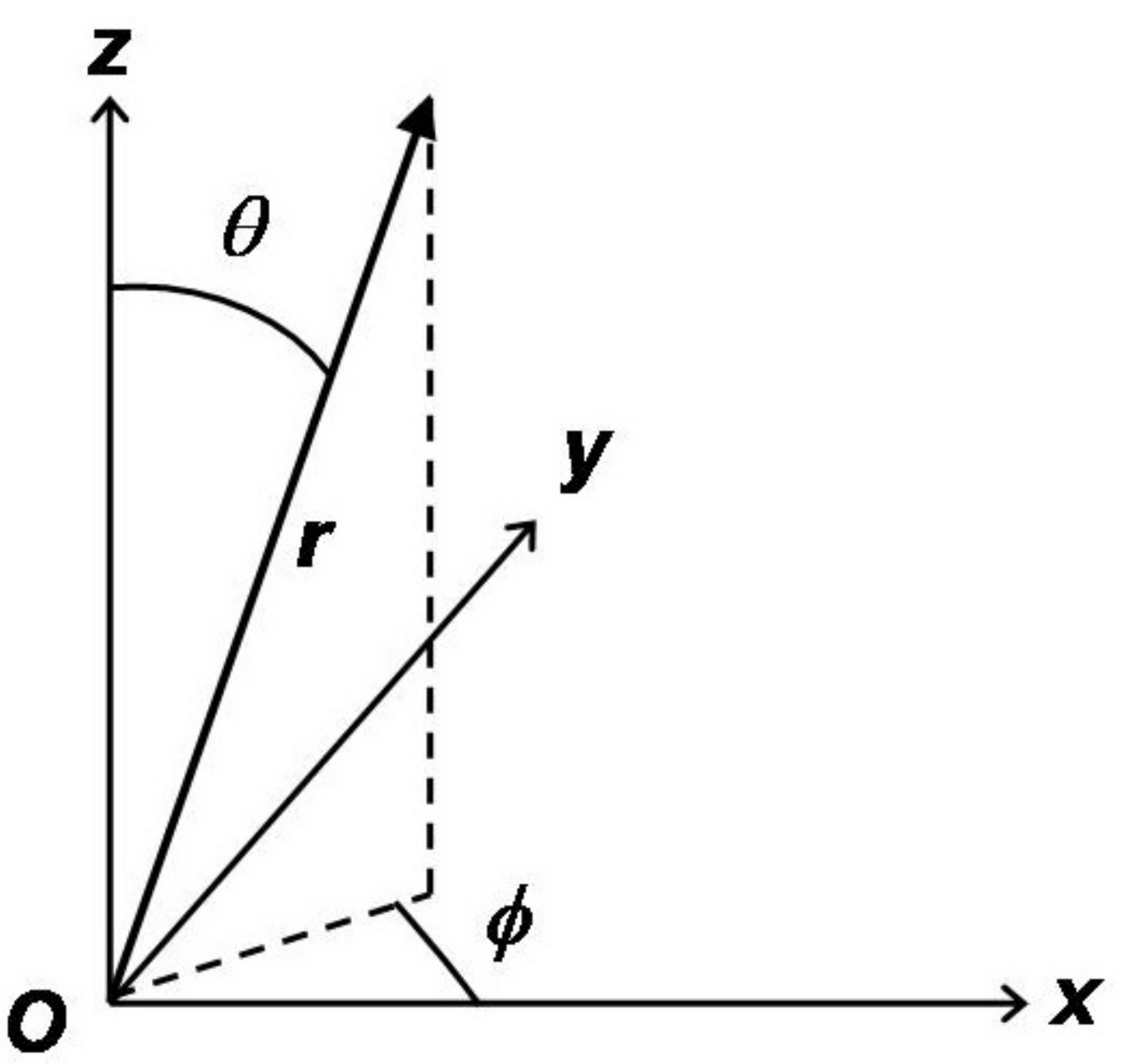}}
\subfigure[]{\label{Fig:Theta sphere}\includegraphics[width=4cm]{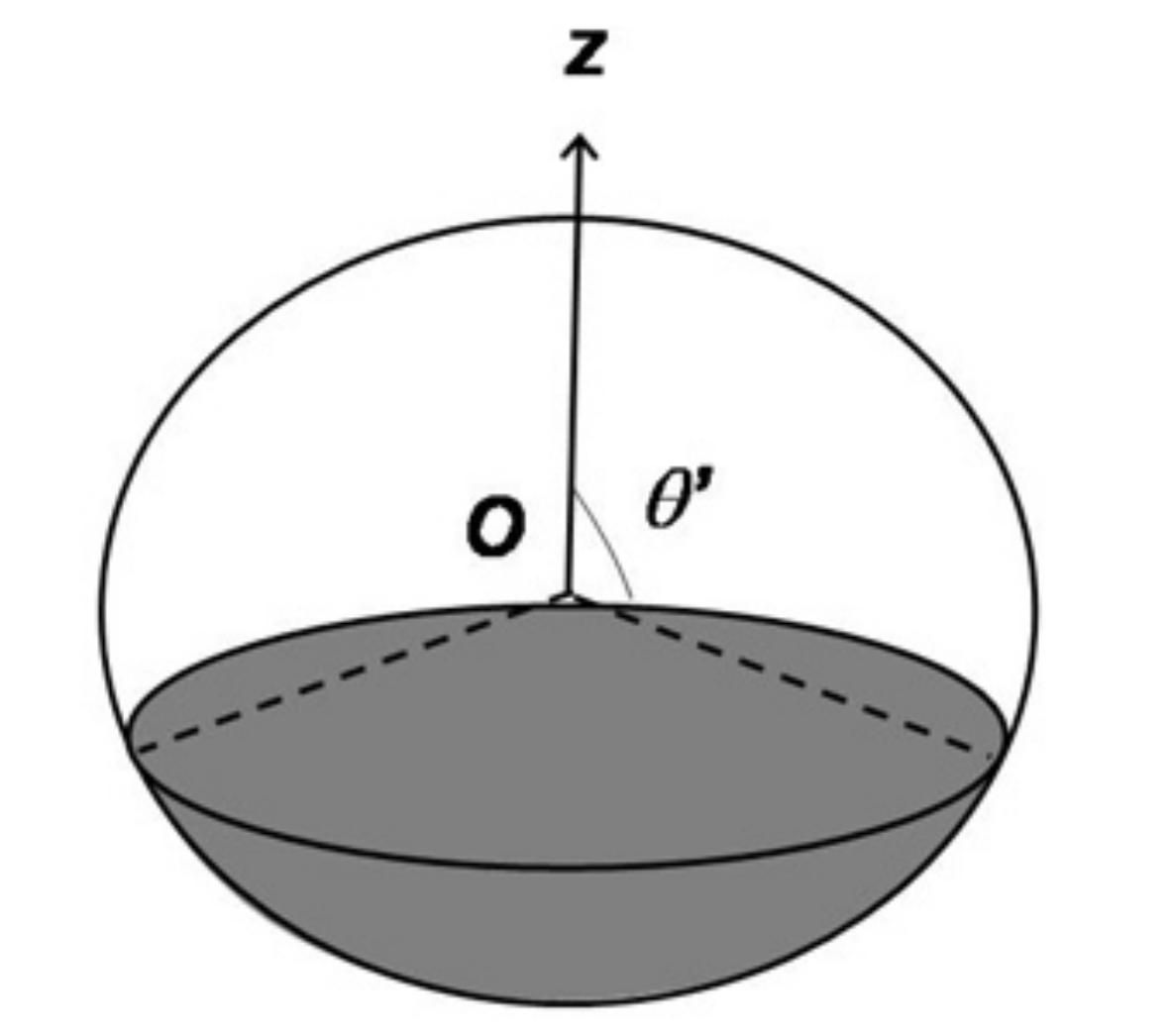}}
\end{center}
\caption{Figure~\ref{Fig:Coordinates} defines Cartesian $\left(x,y,z\right)$ and spherical polar $\left(r,\theta,\phi\right)$ coordinate axes. The centre of the solid sphere of radius $a$ considered in the analyses is located at the origin. Figure~\ref{Fig:Theta sphere} shows a specific inhomogeneous slip configuration (Eq.~\ref{eq:halfBC2}).
} \label{Fig:Axes}
\end{figure}

For an explicit calculation, consider the specific inhomogeneous boundary conditions given by

\begin{equation}\label{eq:halfBC2}
f\left(\theta,\phi\right)=\left\{\begin{array}{cl}
\gamma &\qquad \left(0<\theta<\theta'\right)\\
1 &\qquad \left(\theta'<\theta<\pi\right).
\end{array}\right.
\end{equation}

\noindent which are shown schematically in Fig.~\ref{Fig:Axes}, along with the Cartesian $\left(x,y,z\right)$ and spherical polar $\left(r,\theta,\phi\right)$ co-ordinate axes. To describe this boundary condition, an expansion of normalized Legendre polynomials $P_n\left(\cos\theta\right)$ 
is used, 

\begin{equation}\label{eq:f_half}
\begin{split}
f\left(\theta,\phi\right)&=f_0+f_1+f_2+...\\
&=\frac{1}{2}\left(1+\cos\theta '+\gamma\left(1-\cos{\theta '}\right)\right)P_0\left(\cos\theta\right)\\
&+\frac{3}{4}\left(\cos^2\theta '-1\right)\left(1-\gamma\right)P_1\left(\cos\theta\right)\\
&+\frac{5}{4}\left(\cos^3\theta '-\cos\theta '\right)\left(1-\gamma\right)P_2\left(\cos\theta\right)+ ... .
\end{split}
\end{equation}

\noindent For the hemispherical boundary condition relevant to Janus particles ($\theta '=90$\textdegree ), using Eqs.~\ref{eq:PertForce}, \ref{eq:PertTorque} and \ref{eq:f_half},

\begin{equation}\label{eq:comp_half1}
\mathbf{F}=6\pi\eta a\mathbf{U}\left(1-\frac{b_\parallel\left(1+\gamma\right)}{2a}\right),
\end{equation}

\begin{equation}\label{eq:comp_half3}
\begin{split}
\mathbf{T}&=-6\pi\eta a^2\frac{b_\parallel}{a}\left(1-\gamma\right)\mathbf{U}\times\nabla\left(\frac{3r}{4}\cos\theta\right).\\
\end{split}
\end{equation}

\noindent The rotational motion reaches equilibrium when the vector $\nabla\left(rf_1\right)$ is parallel to $\mathbf{U}$. In Cartesian coordinates, the torque about the centre of the sphere reflects the symmetry of the boundary condition about the $z$~axis:

\begin{equation}\label{eq:comp_half3a}
\begin{split}
\mathbf{T}&=-\frac{9}{2}\pi\eta a^2\frac{b_\parallel}{a}\left(1-\gamma\right)\left(U_y\mathbf{\hat{x}}-U_x\mathbf{\hat{y}}\right).
\end{split}
\end{equation}

The dynamic response of a slipping sphere provides a mechanism for manipulation of that sphere when there is fluid motion relative to the particle. The sphere is lubricated, while asymmetry of the boundary condition relative to the flow direction allows manipulation of orientation. Such dynamics are important, for example, when the particle is tethered stationary relative to flow (but free to rotate), or competing with other applied forces. Application of external fields is not required. The effect of slip increases with $(b$/$a)$, and it is important to recognise that the slip length is not in principle restricted by particle size. Practical manipulation of spheres using asymmetric slip will be dependent on controlling the shape of the particle and the method of inducing slip. Superhydrophobic surfaces generally have physical structure of $O(1)$~$\mu$m while providing Newtonian slip lengths up to $O(10)$~$\mu$m, while slip lengths for water on smooth, hydrophobic surfaces are of $O(10)$~nm. Development of reliable experimental methods for slip length measurement is an area of active research \cite{303}. The major methods currently used do not incorporate significant surface curvature, nor do they precisely test the linearity of Navier's slip length. The calculations described here provide a method for measuring $b$.  

\begin{figure}
\begin{center}
\includegraphics[width=8.5cm]{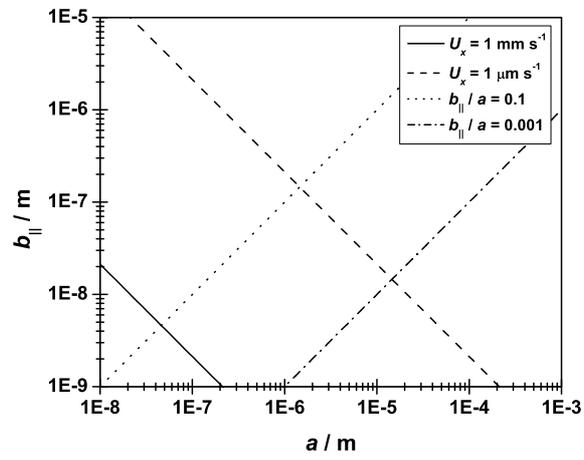}
\end{center}
\caption{Plots of the locus of points at which Pe $=1$ (Eq.~\ref{eq:Energy}). For each line, thermal motion is significant at points closer to the $x$ axis. The fluid is assumed to be water at 300~K.   
} \label{Fig:Peclet}
\end{figure}

The influence of thermal motion on orientation effects can be estimated by considering the Peclet number for the rotating sphere. The mechanical energy scale is the work done by flow parallel to the $x$ axis ($U_x$) when rotating an inhomogeneous slip boundary condition ($\gamma=1$) sphere 90\textdegree\space about the $y$ axis, so that the Peclet number is

\begin{equation}\label{eq:Energy}
\textnormal{Pe}=\frac{9\pi\eta a b_\parallel U_x}{2k_BT},
\end{equation}

\noindent where $T$ is temperature and $k_B$ is Boltzmann's constant. Figure~\ref{Fig:Peclet} is an indicative plot of the conditions at which thermal motion becomes significant (Pe $\approx1$) for water at room temperature. 

To summarize, first-order solutions for low-Reynolds number sphere dynamics that are general with respect to boundary condition inhomogeneity and the relative orientation of unbounded uniform flow have been derived. The full resistance tensors, inclusive of rotational motion, have been presented. The slip perturbation is mathematically equivalent to a slightly deformed sphere, but the governing physical processes for these situations are significantly different. The calculations could be applied to novel applications utilizing inhomogeneous slip. They are just as likely to be useful for characterising the intrinsic properties of particles that are already in use. The analysis will be useful for those interested in slip length measurement, especially for smooth surfaces. Thermal fluctuations will play a significant limiting role in the usefulness of slip-based orientation techniques, especially in atmospheric conditions on the nanoscale.

\clearpage

\clearpage

\end{document}